\documentclass[aps,twocolumn]{revtex4}%
\usepackage{bbm}
\usepackage{amsfonts}
\usepackage{amsmath}
\usepackage{amssymb}
\usepackage{graphicx}%
\providecommand{\U}[1]{\protect\rule{.1in}{.1in}}

\begin{document}
\preprint{HEP/123-qed}
\title[Short title for running header]{Dynamically stabilized decoherence-free states in non-Markovian open fermionic systems}
\author{Heng-Na Xiong}
\affiliation{Department of Physics and Cenert for Quantum information Science, National
Cheng Kung University, Tainan 70101, Taiwan}
\author{Wei-Min Zhang}
\email{wzhang@mail.ncku.edu.tw}
\affiliation{Department of Physics and Cenert for Quantum information Science, National
Cheng Kung University, Tainan 70101, Taiwan}
\author{Matisse Wei-Yuan Tu}
\affiliation{Department of Physics and Cenert for Quantum information Science, National
Cheng Kung University, Tainan 70101, Taiwan}
\author{Daniel Braun}
\email{braun@irsamc.ups-tlse.fr}
\affiliation{Laboratoire de Physique Theorique, Universite Paul Sabatier, Toulouse III, 118
route de Narbonne, F-31062 Toulouse, France}
\date{\today}

\begin{abstract}
Decoherence-free subspaces (DFSs) provide a strategy for protecting the
dynamics of an open system from decoherence induced by the system-environment
interaction. So far, DFSs have been primarily studied in the framework of
Markovian master equations. In this work, we study decoherence-free (DF)
states in the general setting of a non-Markovian fermionic environment. We
identify the
DF states by diagonalizing the non-unitary evolution operator
for a two-level fermionic system attached to an electron reservoir.
By solving the exact master equation, we
show that DF states can be stabilized dynamically.

\end{abstract}
\maketitle

\section{Introduction}

Quantum decoherence is a fundamental issue in open quantum systems and is also
the most important obstacle opposing the realization of a large scale quantum
computer. Unavoidable interactions with noisy environments typically render
initially prepared pure states mixed very rapidly. Nonetheless, a subspace of
Hilbert space can exist where a system undergoes a unitary evolution
irrespective of the interaction with its environment. Such a decoherence-free
subspace (DFS) can in principle provide a theoretically perfect strategy to
protect a system against quantum decoherence. The possibility of DFSs was
already pointed out by Zurek \cite{Zurek82}, who observed that if the
interaction Hamiltonian of the system with the environment has a degenerate
eigenvalue, then superpositions of the corresponding eigenstates remain
coherent. Generally, degenerate eigenvalues can notably arise if the
coupling has a certain symmetry, and the resulting protection against
decoherence has been observed early in the context of rotational tunneling
\cite{Wilson35,Longuet-Higgins63,Freed65,Stevens83,Haeusler85,Wuerger89,Braun93,Braun94}.
In the late 1990s, DFSs were rediscovered independently by several groups
\cite{Palma96,Zanardi97a,Zanardi97b,Duan97,Lidar98,Braun98}. The theory of DFSs
has already been extensively discussed in the context of the
symmetries of the Hamiltonian \cite{Zanardi97a,Zanardi97b}, semigroup dynamics
in the language of quantum master equation
\cite{Lidar98,Braun98,Zanardi98,Lidar05,Lidar99b,Karasik08}, and the operator
sum representation based on Kraus operators \cite{Lidar99a}. Meanwhile, the existence of
DFSs has also been verified experimentally with polarization-entangled photons
\cite{Kwiat00,Altepeter04,Yamamoto08}, trapped ions
\cite{Kielpinski01,Langer05}, nuclear spins using nuclear magnetic resonance
techniques \cite{Viola01,Fortunato02,Ollerenshaw03,Wei05}, and neutron
interferometry \cite{Pushin11}. These experiments show that encoding quantum
information in a DFS can significantly prolong the storage time. Therefore,
DFS have attracted wide interest for applications in fault-tolerant quantum
computation \cite{Lidar99b,Bacon00,Beige00,Beige00b,Zhou04,Wu05,Xue06,Monz09},
long distance quantum communication \cite{Xue08}, quantum key distribution
\cite{Yin08}, quantum teleportation \cite{Wei07}, quantum metrology
\cite{Roos06}, robust quantum repeaters \cite{Klein06,Klein08}, and coherent
quantum control \cite{Cappellaro07,Kiffner07,Wang10}. The requirement of a
symmetry in the coupling to the environment is not always
necessary \cite{Braun00}. Recently, a new Heisenberg-limited
metrology protocol was proposed which exploits the evolution of a DFS due to
a collective
change of the couplings to the environment  \cite{Braun11}.

According to Refs.~\cite{Lidar98,Zanardi98}, a DFS in a system ruled by
Markovian dynamics is defined formally by the vanishing of the
nonunitary Lindblad (decohering) part in the Markovian master equation
\cite{Lindblad76}, see Eq.~(\ref{criterion-0-for-DFS}). This defines a
Hilbert subspace in which dynamics is locally (in time) unitary.
Even so, decoherence can still arise if the Hamiltonian of the system drives a state
out of the DFS. This happens on the time scale of the system Hamiltonian
that is typically much longer than the microscopic decoherence times \cite{Strunz01}.
However, leaking out of the DFS can be
suppressed by coupling the system relatively strongly to the environment
\cite{Beige00,Ognyan10}. An alternative definition of a DFS in
the Markovian context has been given in \cite{Karasik08}. These authors not
 only derived  necessary and
sufficient conditions for the vanishing of the nonunitary part of the master
equation [see Eq.~(\ref{criterion-1-for-DFS})], but also found criteria for
globally DF states, for which the dynamics resulting from the full master
equation (including the Hamiltonian part) remains unitary.

The extension of such DFS criteria to non-Markovian dynamics is not straightforward.
The general non-Markovian master equation of the Nakajima-Zwanzig form
\cite{N58Z60} involves a complicated time-non-local memory integrand
in the nonunitary terms. However, the exact
master equations that describe the general non-Markovian dynamics have been
recently developed for some class of open quantum systems,
including quantum Brownian motion \cite{Hu92,Karrlein97,Kaake85},
entangled cavities with vacuum fluctuations \cite{AnZhang07},
coupled harmonic oscillators \cite{Chou08, Paz0809}, quantum dot
electronic systems in nanosturctures \cite{TuZhang08,TuZhang09}, various nano
devices with time-dependent external control fields \cite{JinZhang10},
nanocavity systems including initial system-reservoir correlations
\cite{TanZhang11}, and photonic networks imbedded in photonic
crystals \cite{LeiZhang12}. These exact master equations can all have
the nonunitary Lindblad form, but the decoherence
rates are time-dependent and may become negative during the time evolution.
This is different from the Markovian case, where the
decoherence rates are always positive.  Since the fact that all rates have
the same sign plays an
essential role in showing the necessity of the DFS criterion,
other possibilities of creating a DF state may arise in non-Markovian cases.

In this work, we study the dynamics of an open fermionic system by solving the
exact non-Markovian fermionic master equation
\cite{TuZhang08,TuZhang09,JinZhang10}. We find a new type of pure DF states
which arise from the fact that certain time-dependent decoherence rates in the
master equation switch themselves off after the system reaches a stable state.
We call these DF states dynamically stabilized DF states.
They are generated in particular for non-Markovian environments. The mechanism
how the DF states arise is therefore very different from the known mechanism
obtained from the Markovian master equation.
For a two-level fermionic system coupled to an electron
reservoir, we find two dynamically stabilized DF states that possess
full quantum coherence between the singly occupied states of the original two
levels. Practical applications of the dynamically stabilized DF states in electron
spin and charge qubits for quantum information processing are expected.

The paper is organized as follows. In Sec.~II, we briefly review the general
criterion of DFSs based on the Markovian master equation formalism. In
Sec.~III, we discuss the dynamics of electron systems in nanostructures via the
exact master equation. In particular, we consider a two-level electron system
whose two levels are coupled identically to an electron reservoir. We find
that the system Hilbert space can be split into two closed subspaces.
Then in Sec.~IV, in terms of the full Lindblad generator, we discuss all the possibilities
how DF states can arise in the system.
We find that the vanishing of one of the two decoherence rates in the
problem can give rise to
dynamically stabilized DF states.  With only one decoherence term in the
master equation remaining, the well-known criterion of DFSs for Markovian
decoherence still provides a necessary and sufficient condition for DF states.
In Sec.~V, we investigate under what conditions the dynamically
stabilized DF states can be reached by the same dissipative process that
allows their existence.  We find that
the initial state and the details of the dynamics given by time-dependent
non-equilibrium Green functions determines which DF state can be reached.
In Sec.~VI, we discuss the generation of DF states in the
Born-Markovian (BM) dynamics within our exact framework. We show that the DF
state in the BM dynamics is a special case of the exact solution. Finally, a
conclusion is given in Sec.~VI.

\section{Criterion for DFS in Markovian case}

In this section, we will briefly review the general criterion of DFS based on
the Markovian master equation following Ref.~\cite{Karasik08}.
Consider an open system $S$ coupled to an external noisy environment $E$. One
can write the total Hamiltonian as
\begin{equation}
H=H_{S}\otimes I_{E}+I_{S}\otimes H_{E}+H_{I},
\end{equation}
where $I$ is the identity operator\ and $H_{I}$ denotes the interaction
Hamiltonian between the system and its environment. If the dynamics of the
system-bath is Markovian and the system and the environment are initially
decoupled, the master equation for the density matrix of the system takes the
Lindbladian form \cite{Carmichael93,Breuer07,Weiss08}
\begin{subequations}
\label{Markov}%
\begin{align}
\dot{\rho}\left(  t\right)   &  =-i\left[  \widetilde{H}_{S},\rho\left(
t\right)  \right]  +L\left[  \rho\left(  t\right)  \right]
,\label{markov-master-eq}\\
L\left[  \rho\right]   &  =\frac{1}{2}\sum_{\alpha=1}^{N}a_{\alpha}\left(
2F_{\alpha}\rho F_{\alpha}^{\dagger}-F_{\alpha}^{\dagger}F_{\alpha}\rho-\rho
F_{\alpha}^{\dagger}F_{\alpha}\right)  , \label{markov-lindblad}%
\end{align}
where $\widetilde{H}_{S}=H_{S}+\Delta$ is the renormalized system Hamiltonian
with $\Delta$ a possible hermitian contribution from the environment ("Lamb
shift"), $F_{\alpha}$ are orthogonal operators on the system Hilbert space
$\mathcal{H}_{S}$, and $a_{\alpha}$ denote real positive coefficients. Thus,
the commutator involving $\widetilde{H}_{S}$ in Eq.~(\ref{markov-master-eq})
determines an effectively unitary evolution of the system, while the decohering
effect induced by the environment is totally accounted for by the non-unitary
term $L\left[  \rho\right]  $.

For the Markovian master equation (\ref{Markov}), the condition of an
instantaneous DFS at time $t$ amounts to the vanishing of $L\left[
\rho\right]  $, that is \cite{Karasik08},
\end{subequations}
\begin{equation}
L\left[  \rho_{\text{DF}}\left(  t\right)  \right]  =0.
\label{criterion-0-for-DFS}%
\end{equation}
This ensures that $\rho_{\text{DF}}\left(  t\right)  $ obeys a unitary
evolution $\dot{\rho}_{\text{DF}}\left(  t\right)  =-i\left[  \widetilde
{H}_{S},\rho_{\text{DF}}\left(  t\right)  \right]  $ at time $t$. This does
not imply necessarily unitary evolution at all times, as the evolution due to
$\tilde{H}_{S}$ can drive the system out of the DFS. In \cite{Lidar98} the DFS
found from (\ref{criterion-0-for-DFS}) was therefore called DFS to the first order.

A sufficient and necessary condition for a state $|k_{\mathrm{DF}}\rangle$ to
be locally DF is given in terms of the operators $\left\{  F_{\alpha}\right\}
$ \cite{Karasik08}
by
\begin{align}
F_{\alpha}|k_{\text{DF}}\rangle &  =c_{\alpha}|k_{\text{DF}}\rangle
,\forall\alpha,k_{\text{DF}}\label{criterion-1-for-DFS}\\
&  \mbox{ and }\sum_{\alpha=1}^{N}a_{\alpha}F_{\alpha}^{\dagger}F_{\alpha
}|k_{\mathrm{DF}}\rangle=\sum_{\alpha=1}^{N}a_{\alpha}|c_{\alpha}%
|^{2}|k_{\mathrm{DF}}\rangle\nonumber
\end{align}
Eq.~(\ref{criterion-1-for-DFS}) means that the DF states\ are degenerate
eigenstates of all the operators $\left\{  F_{\alpha}\right\}  $, and
degenerate eigenstates of $\sum_{\alpha=1}^{N}a_{\alpha}F_{\alpha}F_{\alpha
}^{\dagger}$. A special case is given by $c_{\alpha}=0$ $\forall\alpha$, in
which case the second condition in (\ref{criterion-1-for-DFS}) is
automatically fulfilled. In \cite{Karasik08} it was shown that the space
spanned by the $|k_{\mathrm{DF}}\rangle$ is a DFS at all times, if and only if
in addition to satisfying (\ref{criterion-1-for-DFS}) it is also invariant
under $\widetilde{H}_{S}$.

The generalization of (\ref{criterion-1-for-DFS}) to non-Markovian
master equations is not straight-forward. However, it was recently shown
\cite{Kossakowski10} that under certain conditions the solution of a
non-Markovian master equation \cite{N58Z60} with a finite time memory
kernel can be at the same time a solution of a local-in-time non-Markovian
master equation. In principle, local in time generalizations of the Markovian master equation
(\ref{markov-lindblad}) may be obtained by making both the rates $a_{\alpha}$
and the operators $F_{\alpha}$ time-dependent.
Interestingly, the exact master equations describing the general non-Markovian dynamics for a
large class of bosonic and fermionic systems
\cite{AnZhang07,TuZhang08,TuZhang09,JinZhang10,TanZhang11,LeiZhang12} have
been developed recently. They all have
the Lindbladian form of Eq.~(\ref{Markov}), but the decoherence rates
$a_\alpha$ in
the nonunitary term (\ref{markov-lindblad}) are time-dependent and local
in time, whereas the operators $F_\alpha$ are time-independent. The
time-dependent decoherence rates  are determined
microscopically and nonperturbatively by the retarded and correlation Green
functions in non-equilibrium Green function theory \cite{Sch61407}, where
the back-actions from reservoirs are fully taken into account. As a result,
the non-Markovian dynamics are fully characterized in terms of the
time-non-local retarded integrand in the Dyson equation, which governs the
nonequilibrium Green functions \cite{Kad62,Mahan00}.
In addition to being time-dependent, the decoherence
rates can become negative for short times, representing in a
certain sense the back-flow of information from the environment to the system
\cite{TuZhang08,Piilo09,XiongZhang10,WuZhang10,LeiZhang11}. Since the
fact that all $a_{\alpha}$ have the same sign is an important requirement for showing that
(\ref{criterion-1-for-DFS}) is necessary for a Markovian DF state  (see the argument
after eq.(3.15) in \cite{Karasik08}),
additional DF states may arise in the
non-Markovian case.
In the following, we will provide a new mechanism for generating
dynamically stabilized DF states in fermionic systems, based on the exact
fermionic master equation developed recently
\cite{TuZhang08,TuZhang09,JinZhang10}.

\section{Exact Master Equation}

We consider a general nanoelectronic system with $N$ energy levels coupled to
an electron reservoir. The Hamiltonian for the system, the electron reservoir,
and the interaction between them read
\begin{align}
H_{S}  &  =\sum_{i=1}^{N}\epsilon_{i}a_{i}^{\dagger}a_{i},~~~H_{B}=\sum
_{k}\mathbf{\varepsilon}_{k}c_{k}^{\dagger}c_{k},\nonumber\\
H_{I}  &  =\sum_{i=1}^{N}\sum_{k}\left(  V_{ik}e^{i\phi_{i}}a_{i}^{\dagger
}c_{k}+V_{ik}e^{-i\phi_{i}}c_{k}^{\dagger}a_{i}\right)  .
\label{general-Hamiltonian}%
\end{align}
Here $a_{i}^{\dagger}$ and $a_{i}$ are electron creation and annihilation
operators for $i$th level with energy $\epsilon_{i}$.
The operators $c_{k}^{\dagger}$ and $c_{k}$ denote electron creation
and annihilation operators for the energy level $\mathbf{\varepsilon}_{k}$ of
the electron reservoir. The coupling strength between the system and the
reservoir is described by the $V_{ik}\in\mathbb{R}$, with the phase $\phi_{i}$ of
the coupling made explicit. The total hamiltonian is $H=H_S+H_B+H_I$.

The exact master equation for such system was obtained in
\cite{TuZhang08,TuZhang09,JinZhang10},
\begin{subequations}
\begin{equation}
\dot{\rho}\left(  t\right)  =-i\left[  \widetilde{H}_{S}\left(  t\right)
,\rho\left(  t\right)  \right]  +L\left[  \rho\left(  t\right)  \right]  ,
\end{equation}
where the renormalized system Hamiltonian $\widetilde{H}_{S}\left(  t\right)
$ and the decoherence term $L\left[  \rho\left(  t\right)  \right]  $ take the
form
\begin{align}
&  \widetilde{H}_{S}(t) =\sum_{i,j=1}^{N}\widetilde{\epsilon}_{ij}(t)
a_{i}^{\dagger}a_{j},\\
L[ \rho(t) ] =\sum_{i,j=1}^{N}\big\{  &  \kappa_{ij}(t) \big[ 2a_{j}%
\rho(t)a_{i}^{\dagger}-a_{i}^{\dagger} a_{j}\rho(t) -\rho(t) a_{i}^{\dagger
}a_{j}\big]\nonumber\\
+ \widetilde{\kappa}_{ij}(t) \big[ 2  &  a_{i}^{\dagger}\rho(t) a_{j}%
-a_{j}a_{i}^{\dagger}\rho(t) -\rho(t) a_{j}a_{i}^{\dagger}\big] \big\} .
\label{exact-ME-Lrho}%
\end{align}
Here the shifted $\widetilde{\epsilon}_{ij}(t) $ and decoherence rates
$\kappa_{ij}(t) $ and $\widetilde{\kappa}_{ij}(t) $ are all time-dependent but
local in time. They are determined microscopically and nonperturbatively in
terms of the retarded and correlation Green functions by eliminating
completely all the reservoir degrees of freedom (i.e. tracing over all the
states of the environment). Their explicit forms are shown in Eq.~(9) of Ref.
\cite{TuZhang08}.

To be more specific, let us consider the case of a nanoelectronic system with
$N=2$. Physically, such a system may be realized by a double quantum dot
system in which each dot has a single active energy (on-site) level, coupled
to electrodes with all spins polarized in both the dots and the electrodes.
Another example for $N=2$ is given by a single-level quantum dot coupled to
electrodes with allowed spin flips between two antiparallel directions.

Furthermore, we assume that the two
energy levels of the system are degenerate: $\epsilon_{1}=\epsilon
_{2}=\epsilon_{0}$. Practically, the energy degeneracy
is easier to be realized  in the second setting than in the first. We also
assume that both levels have the same
coupling strength to the electron reservoir, i.e., $V_{1k}=V_{2k}=V_{k}%
/\sqrt{2}$. Then we introduce two effective fermion operators
\end{subequations}
\[
A_{+}=\frac{1}{\sqrt{2}}\left(  a_{1}+e^{i\phi}a_{2}\right)  ,~A_{-}=\frac
{1}{\sqrt{2}}\left(  -e^{-i\phi}a_{1}+a_{2}\right)  ,
\]
with $\phi=\phi_{1}-\phi_{2}$. The Hamiltonian (\ref{general-Hamiltonian}) can
be rewritten in terms of $A_{\pm}$ as follows
\begin{align}
H_{S}  &  =\epsilon_{+}A_{+}^{\dagger}A_{+}+\epsilon_{-}A_{-}^{\dagger}
A_{-},H_{B}=\sum_{k}\varepsilon_{k}c_{k}^{\dagger}c_{k},\nonumber\\
H_{I}  &  =\sum_{k}V_{k}\left[  e^{i\phi_{1}}A_{+}^{\dagger}c_{k}
+e^{-i\phi_{1}}c_{k}^{\dagger}A_{+}\right]  . \label{Hamiltonian-two-dots}
\end{align}
where the effective energy levels $\epsilon_{\pm}=\epsilon_{0}$ are still degenerate.

The system Hamiltonian is diagonalized in terms of $A_{\pm}$,
such that the original system is equivalent to an effective system
which has two decoupled energy levels $\epsilon_{\pm}$, out of which only one
energy level ($\epsilon_{+}$) couples to the electron reservoir.
As a result, the corresponding exact master equation becomes
\begin{subequations}
\label{exact-master-equation}%
\begin{equation}
\dot{\rho}\left(  t\right)  =-i\left[  \widetilde{H}_{S}\left(  t\right)
,\rho\left(  t\right)  \right]  +L\left[  \rho\left(  t\right)  \right]  ,
\end{equation}
with the new expressions of $\widetilde{H}_{S}\left(  t\right)  $ and
$L\left[  \rho\left(  t\right)  \right]  $
\begin{align}
\label{exact-master-equation-H-L} &  ~~~~~~~~~~~\widetilde{H}_{S}\left(  t\right)
=\widetilde{\epsilon}_{+}\left(  t\right)  A_{+}^{\dagger}A_{+}+\epsilon
_{-}A_{-}^{\dagger}A_{-},\\
&  L\left[  \rho\left(  t\right)  \right]  =\kappa\left(  t\right)  \left[
2A_{+}\rho\left(  t\right)  A_{+}^{\dagger}-A_{+}^{\dagger}A_{+}\rho\left(
t\right)  -\rho\left(  t\right)  A_{+}^{\dagger}A_{+}\right] \nonumber\\
&  ~~~+\widetilde{\kappa}\left(  t\right)  \left[  2A_{+}^{\dagger}\rho\left(
t\right)  A_{+}-A_{+}A_{+}^{\dagger}\rho\left(  t\right)  -\rho\left(
t\right)  A_{+}A_{+}^{\dagger}\right]  . \label{decohering}%
\end{align}
The decoherence rates are $\kappa\left(  t\right)  =\gamma\left(  t\right)
-\frac{\widetilde{\gamma}\left(  t\right)  }{2}$ and $\widetilde{\kappa
}\left(  t\right)  =\frac{\widetilde{\gamma}\left(  t\right)  }{2}$. The
renormalized energy level $\widetilde{\epsilon}_{+}\left(  t\right)  $ and the
rates $\gamma\left(  t\right)  $ and $\widetilde{\gamma}\left(
t\right)  $ are determined exactly by
\end{subequations}
\begin{subequations}
\label{decohering-coefficients}%
\begin{align}
&  \widetilde{\epsilon}_{+}(t) =-\mathrm{Im}[ \dot{u}(t) u^{-1}(t)] ,\\
&  \gamma(t) =- \mathrm{Re}[ \dot{u}(t) u^{-1}(t)] ,\\
&  \widetilde{\gamma}(t) =\dot{v}(t) -2v(t) \mathrm{Re}[ \dot{u}(t) u^{-1}(t)]
,
\end{align}
and $u(t)$ and $v\left(  t\right)  $ are the retarded and correlation Green
functions in the Schwinger-Keldysh nonequilibrium Green function theory
\cite{JinZhang10}. They obey the integral-differential equations
\end{subequations}
\begin{subequations}
\label{uv-eq}%
\begin{align}
&  \frac{d}{dt}u\left(  t\right)  +i\epsilon_{+}u\left(  t\right)
+\int_{t_{0}}^{t}g\left(  t-\tau\right)  u\left(  \tau\right)  d\tau
=0,\label{u-eq}\\
&  v\left(  t\right)  =\int_{t_{0}}^{t}d\tau_{1}\int_{t_{0}}^{t}d\tau
_{2}u\left(  \tau_{1}\right)  \widetilde{g}\left(  \tau_{2}-\tau_{1}\right)
u^{\ast}\left(  \tau_{2}\right)  , \label{v-eq}%
\end{align}
subject to the boundary conditions $u\left(  t_{0}\right)  =1$ . The
integration kernels read $g\left(  \tau\right)  =\int_{-\infty}^{+\infty}%
\frac{d\omega}{2\pi}J\left(  \omega\right)  e^{-i\omega\tau}$ and
$\widetilde{g}\left(  \tau\right)  =\int_{-\infty}^{+\infty}\frac{d\omega
}{2\pi}J\left(  \omega\right)  f\left(  \omega\right)  e^{-i\omega\tau}$,
where the spectral density of the reservoir is given by $J\left(
\omega\right)  =2\pi\sum_{k}\left\vert V_{k}\right\vert ^{2}\delta\left(
\omega-\omega_{k}\right)  $, and $f\left(  \omega\right)  =1/\left(
e^{\beta\left(  \omega-\mu\right)  }+1\right)  $ is the initial electron
distribution of the reservoir.

Since the effective energy level $\epsilon_{-}$ is apparently decoupled from
the electron reservoir, one may naively think that the state generated by the
operator $A^{\dag}_{-}$, namely $A^{\dag}_{-}|0\rangle$ becomes naturally a DF
state. This is actually not true. The term with $\widetilde{\kappa}(t)$ in
(\ref{decohering}) will drive this state into another state since $A^{\dag
}_{+}A^{\dag}_{-}|0\rangle\neq0$. However, there is an occupation constant of
motion in this system, $\left[  A_{-}^{\dagger}A_{-},H\right]  =0$. This
symmetry separates the system Hilbert space into two closed subspaces
$\mathcal{H}_{+}=\left\{  |\mathsf{v}\rangle,|+\rangle\right\}  $ and
$\mathcal{H}_{-}=\left\{  |-\rangle,|\mathsf{d}\rangle\right\}  $,
corresponding to the occupation $N_{-} \equiv\langle A_{-}^{\dagger}A_{-}
\rangle=0$ and $1$, respectively. Here $|\mathsf{v}\rangle$ ($|\mathsf{d}%
\rangle$) is the vacuum (doubly occupied) electron state, while $|\pm
\rangle=A_{\pm}^{\dagger}|\mathsf{v}\rangle$ are superpositions of singly
occupied states of the original two coupled levels,
\end{subequations}
\begin{align}
|+\rangle &  =\left(  |1\rangle+e^{-i\phi}|2\rangle\right)  /\sqrt
{2},\nonumber\\
|-\rangle &  =\left(  -e^{i\phi}|1\rangle+|2\rangle\right)  /\sqrt{2},
\label{scs}%
\end{align}
and $|i\rangle=a_{i}^{\dagger}|\mathsf{v}\rangle$ ($i=1,2$). The two states
$|\pm\rangle$ are the superpositions of the two original energy levels with
the relative phases $\phi$ and $\phi+\pi$, where $\phi$ is arbitrary. As a
consequence, starting from any initial state in the closed subspace
$\mathcal{H}_{+}$ ($\mathcal{H}_{-}$), the system will be kept in this
subspace throughout the evolution process.

Eq.(\ref{exact-master-equation}) has the standard form of Lindblad
master equation, except that the decoherence rates, $\kappa(t)$ and
$\widetilde{\kappa}(t)$, can depend on time and even become negative. They
are local-in-time and determined
microscopically and nonperturbatively from Eq.~(\ref{uv-eq}). The
DFS criterion of Eq.~(\ref{criterion-1-for-DFS}) is then still sufficient,
but may not be necessary. The operators $A_{+}^{\dag}$ and $A_{+}$ act on
the four basis
states $\{|$v$\rangle,|-\rangle,|+\rangle,|$d$\rangle\}$ according to the
following relations:
\[%
\begin{array}
[c]{lll}%
A_{+}^{\dagger}|\mathsf{v}\rangle=|+\rangle, & ~~ & A_{+}|\mathsf{v}%
\rangle=0,\\
A_{+}^{\dagger}|+\rangle=0, &  & A_{+}|+\rangle=|\mathsf{v}\rangle,\\
A_{+}^{\dagger}|-\rangle=|\mathsf{d}\rangle, &  & A_{+}|-\rangle=0,\\
A_{+}^{\dagger}|\mathsf{d}\rangle=0, &  & A_{+}|\mathsf{d}\rangle=|-\rangle.
\end{array}
\]
One checks that both $A_{+}^{\dagger}$ and $A_{+}$ have a four-fold degenerate
eigenvalue $c_{\alpha}=0$, but the corresponding eigenspaces are only
two-dimensional, given by $\{|+\rangle,|$d$\rangle\}$ for $A_{+}^{\dagger}$
and $\{|-\rangle,|$v$\rangle\}$ for $A_{+}$. Since these two spaces do not
overlap, condition (\ref{criterion-1-for-DFS}) is not satisfied, and as
long as both $\kappa(t)$ and $\widetilde{\kappa}(t)$ are positive,
there is no DFS. In the next section, we explore whether additional DF states can exist if
$\kappa(t)$ and $\widetilde{\kappa}(t)$ are not both positive, and show
that a DFS can arise dynamically,
when at least one of the two rates $\kappa(t)$ and $\widetilde{\kappa}(t)$
vanishes.

\section{DF states}

\label{chap-IV}
\subsection{Eigenstates of the Lindblad operator}
Since, as discussed above, the Markovian DFS criterion
(\ref{criterion-1-for-DFS}) may not be necessary for a
non-Markovian system, we define a local (in time) DFS
by the vanishing of the non-unitary term $L[\rho(t)]$. Clearly, this leads to
local unitary time-evolution.
In the following, we will discuss all possibilities how $L[\rho(t)]=0$
can arise by calculating the eigenvalues of
the full Lindblad generator $L $ in
Liouville space. If $L$ has a zero eigenvalue,
the corresponding eigenstate is
DF. Conversely, a local DF state is by definition an eigenstate of $L$ with
eigenvalue zero.  However, as the
eigenstates of
$L$ do not necessarily  have all the properties of a density
matrix, such states may not be physical. One must therefore examine for each
eigenstate whether it is a physical state or can be combined with
other eigenstates corresponding to the same (degenerate) eigenvalue to form a
physical state. In this way one can find all physical DF
states possible.

In the basis of $\left\{  |\mathsf{v}\rangle
,|+\rangle,|-\rangle,|\mathsf{d}\rangle\right\}  $, the density matrix
takes the general form
\begin{align}
\rho\left(  t\right)  =  &  \rho_{\mathsf{v v}}\left(  t\right)
|\mathsf{v}\rangle\langle\mathsf{v}| +\rho_{++}\left(  t\right)
|+\rangle\langle+|\nonumber\\
&  +\rho_{+-}(t) |+\rangle\langle-|+\rho_{+-}^{\ast}\left(  t\right)
|-\rangle\langle+|\nonumber\\
&  +\rho_{--}\left(  t\right)  |-\rangle\langle-|+\rho_{\mathsf{d d}}\left(
t\right)  |\mathsf{d}\rangle\langle\mathsf{d}| . \label{general-rhot}%
\end{align}
Coherences between states with different particle numbers are not permitted
due to the particle number super-selection rule.
Since there are only six nonzero elements of $\rho\left(  t\right)  $,
we can define the
basis $\{|1\rangle\equiv|$%
v$\rangle\langle$v$|$, $|2\rangle\equiv|+\rangle\langle+|$, $|3\rangle
\equiv|+\rangle\langle-|$, $|4\rangle\equiv|-\rangle\langle+|$, $|5\rangle
\equiv|-\rangle\langle-|$,
$|6\rangle\equiv|$d$\rangle\langle$d$|\}$ in Liouville space, which
is orthogonal with
respect to the scalar product $\langle A|B\rangle={\rm tr}{A^\dagger B}$.
In this basis, the density matrix can be rewritten as a column vector,
$|\rho\left(  t\right)
\rangle=\left[  \rho_{\text{vv}}\left(  t\right)  ,\rho_{++}\left(  t\right)
,\rho_{+-}\left(  t\right)  ,\rho_{-+}\left(  t\right)  ,\rho_{--}\left(
t\right)  ,\rho_{\text{dd}}\left(  t\right)  \right]  ^{T}$. The
Lindblad operator $L$ is represented by a matrix $L_t$, and its action
on a density matrix reduces to a simple matrix multiplication of $L_t$
with $|\rho(t)\rangle$,
where
$L_{t}$ reads \begin{widetext}
\begin{equation}
L_{t}=\left[
\begin{array}
[c]{cccccc}%
-2\widetilde{\kappa}\left(  t\right)   & 2\kappa\left(  t\right)   & 0 & 0 &
0 & 0\\
2\widetilde{\kappa}\left(  t\right)   & -2\kappa\left(  t\right)   & 0 & 0 &
0 & 0\\
0 & 0 & -\left(  \kappa\left(  t\right)  +\widetilde{\kappa}\left(  t\right)
\right)   & 0 & 0 & 0\\
0 & 0 & 0 & -\left(  \kappa\left(  t\right)  +\widetilde{\kappa}\left(
t\right)  \right)   & 0 & 0\\
0 & 0 & 0 & 0 & -2\widetilde{\kappa}\left(  t\right)   & 2\kappa\left(
t\right)  \\
0 & 0 & 0 & 0 & 2\widetilde{\kappa}\left(  t\right)   & -2\kappa\left(
t\right)
\end{array}
\right]  .
\end{equation}
\end{widetext} The eigenvalues and the corresponding eigenstates of
$L_{t}$ are
\begin{align}
l_{1}  &  =0,\text{ \ \ \ \ \ \ \ \ }\nonumber\\
l_{2}  &  =-2\left[  \kappa\left(  t\right)  +\widetilde{\kappa}\left(
t\right)  \right]  ,\nonumber\\
l_{3}  &  =-\left[  \kappa\left(  t\right)  +\widetilde{\kappa}\left(
t\right)  \right]  ,\text{ }\nonumber\\
l_{4}  &  =-\left[  \kappa\left(  t\right)  +\widetilde{\kappa}\left(
t\right)  \right]  ,\text{ }\nonumber\\
l_{5}  &  =-2\left[  \kappa\left(  t\right)  +\widetilde{\kappa}\left(
t\right)  \right]  ,\nonumber\\
l_{6}  &  =0,
\end{align}
and
\begin{align}
|l_{1}\rangle &  =\frac{1}{{\kappa\left(  t\right)  +\widetilde{\kappa}\left(
t\right)  }}\left[  \kappa\left(  t\right)  |1\rangle+\widetilde{\kappa
}\left(  t\right)  |2\rangle\right]  ,\nonumber\\
|l_{2}\rangle &  =\frac{1}{{2}}\left[  -|1\rangle+|2\rangle\right]
,\nonumber\\
|l_{3}\rangle &  =|3\rangle,\nonumber\\
|l_{4}\rangle &  =|4\rangle,\nonumber\\
|l_{5}\rangle &  =\frac{1}{{2}}\left[  -|5\rangle+|6\rangle\right]
,\nonumber\\
|l_{6}\rangle &  =\frac{1}{{\kappa\left(  t\right)  +\widetilde{\kappa}\left(
t\right)  }}\left[  \kappa\left(  t\right)  |5\rangle+\widetilde{\kappa
}\left(  t\right)  |6\rangle\right]  . \label{eigvec-01}%
\end{align}
We see that all eigenvalues come in pairs, and there are always at
least two eigenvalues equal to zero. In writing (\ref{eigvec-01}) we
have assumed that $\kappa(t)+\widetilde{\kappa}(t)\ne 0$. The case of
$\kappa(t)+\widetilde{\kappa}(t)=0$  will be discussed below.  The normalization
used for $|l_1\rangle$ and $l_6\rangle$ is
convenient, as in this way these two states can be interpreted
directly as density matrices if both $\kappa(t)$ and $\widetilde{\kappa}(t)$ are
positive. If one of the rates is negative (and the other non-zero), both
states become non-positive
and therefore cease to be physical states. Moreover, since they have
orthogonal support, no linear combination of them can bring about a
positive state.  Similarly, states $|l_2\rangle$ and $|l_5\rangle$, as
well as any linear combination of them, are
clearly non-positive.  Finally, states $|l_3\rangle$ and $|l_4\rangle$ as
well as any linear combination of them are
traceless and are therefore not physical states either. Since
$|l_2\rangle,\ldots,|l_5\rangle$ are independent of $\kappa(t)$ and
$\widetilde{\kappa}(t)$ (and are therefore {\em never} physical states,
regardless of the values of $\kappa(t)$ and
$\widetilde{\kappa}(t)$), the
only possibility of having a physical DF state is through
$|l_1\rangle$ and $|l_6\rangle$ with both $\kappa(t)$ and
$\widetilde{\kappa}(t)$ non-negative, or one of them vanishing (in the
latter case one may always choose the eigenvector with positive global
sign).

If $\kappa\left(  t\right)  =0$ and $\widetilde{\kappa
}\left(  t\right)  \neq0$, $|l_{1}\rangle$ and $|l_{6}\rangle$
are two pure
DF states $|2\rangle$ and $|6\rangle$, that is, the subset of $\left\{  |+\rangle
,|\text{d}\rangle\right\}$ contains all the possible DF states in this
case. Likewise, if $\widetilde{\kappa
}\left(  t\right)  =0$ and $\kappa\left(  t\right)  \neq0$,
$|l_{1}\rangle$ and $|l_{6}\rangle$
are the two pure DF
states $|1\rangle$ and
$|5\rangle$, i.e., $\left\{  |\text{v}\rangle,|+\rangle\right\}  $.
Below, by examining an explicit example,
we will show that the vanishing of one of the decoherence rates is
physically feasible after some time
$t_s$, when the system reaches its steady state,
see  Fig.~\ref{fig_kappa}.

If $\kappa\left(  t\right)  >0$ and $\widetilde{\kappa
 }\left(  t\right)  >0$, the two eigenstates
 $|l_{1}\rangle$ and
$|l_{6}\rangle$ are time-dependent mixed states. They are decoherence
free as much as they are locally stationary states due to local detailed
balance. From a perspective of application for quantum
information processing, these states are less interesting.  They are
 the analogues of
thermal equilibrium states that are stationary under a Markovian
relaxation process. \\

It remains to consider the case $\kappa(t)+\widetilde{\kappa}(t)=0$. All
eigenvalues vanish, but as long as $\kappa(t)\ne 0$ the
eigenvectors are still given by Eq.~(\ref{eigvec-01}), with the only
difference
that the normalization of $|l_1\rangle$ and $|l_6\rangle$
through the
prefactor $1/(\kappa(t)+\widetilde{\kappa}(t))$ has to be
removed.
$|l_1\rangle$
becomes colinear with $|l_2\rangle$, and $|l_6\rangle$ colinear with
$|l_5\rangle$.  The dimension of the space of eigenvectors of $L_t$ is
reduced to four, and $L_t$ can therefore not be fully diagonalized.
As the linearly independent eigenvectors
$|l_1\rangle,\ldots,|l_4\rangle$ are never physical states, this means
that $L_t$ has no eigenstates that are physically possible, and
therefore no DF states exist.

If $\kappa(t)=\widetilde{\kappa}(t)=0$ at
some time (for
example, if the retarded Green function
$u\left(  t\right)  $ took a nonzero steady value, it would be
possible that both rates vanish for $t>t_s$),  we have $L_t=0$, and
the whole Hilbert space becomes a
dynamically stabilized DFS.  From a quantum information perspective this
 would be of course an ideal situation.  Unfortunately, in the
fermionic system interacting with an electron reservoir considered
here, it appears that this
situation does not arise, as shown in Fig. \ref{fig_kappa}.

In summary, as long as we restrict ourselves to pure states for our
non-Markovian master equation and discard the ``trivial'' case
$L_t=0$, the DFS is
still given entirely by the Markovian criterion
(\ref{criterion-1-for-DFS}), as pure DF states only exist if exactly one of
the two terms (proportional to either $\kappa(t)$ or $\widetilde{\kappa}(t)$)
in the Lindblad superoperator remains, and  the logic
of the proof of necessity of condition  (\ref{criterion-1-for-DFS})
remains intact in such a situation. However, the time dependence of $\kappa(t)$
and $\widetilde{\kappa}(t)$ brings about a new freedom, and allows for the
dynamical stabilization of DF states through the switching off of one of the
decoherence rates.

\subsection{Physical realization of DF states}
In the following, we will discuss to what extent the DF states
just discussed can be reached through the same non-Markovian dynamics
described by $L_t$. We consider the double quantum dot system from
above coupled to an electron reservoir with a spectral density
of the Lorentz form
\cite{Mac06,Jin08,TuZhang08},
\begin{equation}
J\left(  \omega\right)  =\frac{\Gamma d^{2}}{\left(  \omega-\epsilon
_{+}\right)  ^{2}+d^{2}}, \label{Lorentz-spectral}%
\end{equation}
where $\Gamma$ is the system-reservoir coupling strength, and $d$
the bandwidth of the effective reservoir spectrum. In the well-known
wide-band limit, i.e., $d \rightarrow
\infty$, the spectral density approximately becomes a constant one, $J(\omega)
\rightarrow\Gamma$. This corresponds to the Markovian limit.

\begin{figure}[ptb]
\begin{center}
\includegraphics[height=4.5cm,width=5.5cm]{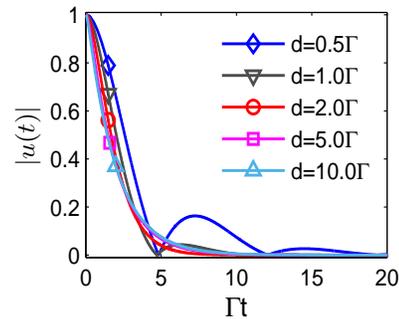}
\end{center}
\caption{The exact solution of $|u(t)|$ for an electron reservoir with the
Lorentz spectral density. Here we take $\epsilon_{0}=0.2\Gamma$. Note that
$|u(t)|$ decays exponentially for large bandwidth $d$, corresponding to the
weakly non-Markovian dynamics. For a small bandwidth $d<2\Gamma$, the strong
non-Markovian memory effect brings the short-time oscillations for $|u(t)|$. }%
\label{fig_ut}%
\end{figure}

With the above spectral density, the solution of $u(t)$ obeying
Eq.~(\ref{u-eq}) can be obtained analytically
\begin{equation}
u\left(  t\right)  =\left\{
\begin{array}
[c]{c}%
\frac{e^{-i\epsilon_{+}t}}{2d_{\Gamma}}\left[  d_{\Gamma}^{+}e^{-\frac
{d_{\Gamma}^{-}t}{2}}-d_{\Gamma}^{-}e^{-\frac{d_{\Gamma}^{+}t}{2}}\right]
,d\neq2\Gamma,\\
\left[  1+\frac{dt}{2}\right]  e^{-\left(  i\epsilon_{+}+\frac{d}{2}\right)
t},\text{ \ \ \ \ \ \ \ \ \ \ }d=2\Gamma,
\end{array}
\right.  \label{u-sol}%
\end{equation}
where $d_{\Gamma}=\sqrt{d^{2}-2\Gamma d}$ and $d_{\Gamma}^{\pm}=d\pm
d_{\Gamma}$. Obviously, after some time $t_{s}$, $u\left(  t\right)  $ always
decays to zero, as shown in Fig. \ref{fig_ut} where the different behaviors of
the amplitude of $u\left(  t\right)  $ corresponds to different bandwidths
$d$. For the bandwidth $d\gtrsim2\Gamma$ (weakly non-Markovian case), $u(t)$
exponentially decays to zero, which is a result similar to the Markovian
dynamics, see the discussion in Sec. VI. When the bandwidth $d<2\Gamma$, the
non-Markovian memory effect of the reservoir becomes significant, which
induces a short-time oscillation for $\left\vert u\left(  t\right)
\right\vert $. 

\begin{figure}[ptb]
\begin{center}
\includegraphics[height=7.3cm,width=9cm]{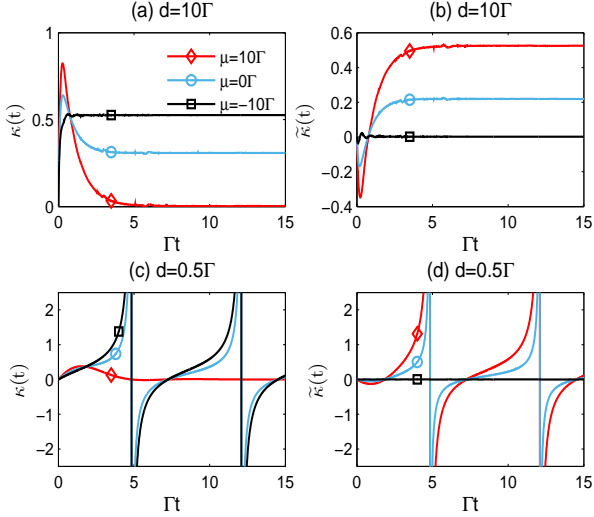}
\end{center}
\caption{The exact solutions of decoherence rates $\kappa(t)$ and
$\widetilde{\kappa}(t)$ for different external bias voltage $\mu=eV$. (a)-(b) for
the weakly non-Markovian dynamics (with $d=10\Gamma$); (c)-(d) for the strong
non-Markovian dynamics (with $d=0.5\Gamma$). Here we set $\epsilon
_{0}=0.2\Gamma$, ${k_{B}}T=0.3\Gamma$. We note that generally, $\kappa(t)$
and $\widetilde{\kappa}(t)$ are non-zero during the time evolution. However, for a
large positive (or negative) bias voltage, one of them is switched off after
some specific time. }%
\label{fig_kappa}%
\end{figure}

Using the solution of Eq.~(\ref{u-sol}), combined with Eqs.~(\ref{v-eq}) and
(\ref{decohering-coefficients}), we can easily calculate the decoherence rates
$\kappa\left(  t\right)  $ and $\widetilde{\kappa}\left(  t\right)  $ in
eq.~(\ref{exact-master-equation}). The result is plotted in
Fig.~\ref{fig_kappa} for the cases of weakly ($d=10\Gamma$) and strongly
($d=0.5\Gamma$) non-Markovian dynamics. In experiments, one can adjust
the external bias ($\mu$)
to raise or lower the Fermi surface of the electron reservoir. Here we
display three cases
where the bias is much higher ($\mu=10\Gamma$), relatively small ($\mu=0$),
and much lower ($\mu=-10\Gamma$)
than the quantum dot energy level $\epsilon_0$. First, we see that
when $d=0.5\Gamma$, as
shown in Fig.~\ref{fig_kappa} (c) and (d), $\kappa\left(  t\right)  $ and
$\widetilde{\kappa}\left(  t\right)  $ can jump from a positive value to
a negative value repeatedly during the evolution process, which corresponds to
the forth- and back-flows of the information between the system and the
environment, in evidence of the strong non-Markovian dynamics.

We observe that, no matter what the values of the width $d$ and
the bias $\mu$ are, the two decoherence rates
$\kappa(t)$ and $\widetilde{\kappa}(t)$ never satisfy
$\kappa(t)+\widetilde{\kappa}(t)=0$.
However, if we apply a large bias to raise (or
lower) the Fermi surface of the electron
reservoir much higher (or much lower) than the dot level, one of the two
decoherence rates is switched off after a time scale of a few
$1/\Gamma$, which implies the existence of a DFS.

As shown in Fig.~\ref{fig_kappa} (a) and (b), in the weakly non-Markovian
case, when applying
a positive bias $\mu=10\Gamma$, the decoherence rate $\kappa\left(  t\right)
$ shows a positive peak in the beginning and then decays rapidly to a zero
steady value on a time scale of a few $1/\Gamma$, while $\widetilde{\kappa
}\left(  t\right)  $ turns negative first, and then climbs to a nonzero steady
value. For the strongly non-Markovian regime ($d=0.5\Gamma$), the
same situation happens, see the red curves in Fig.~\ref{fig_kappa} (c) and
(d). The decoherence rate reaches $\kappa(t)=0$ on a time scale $t_s$ of a
few
$1/\Gamma$, while $\widetilde{\kappa}(t)$ keeps jumping from positive values
to negative values repetitively. In this case, the dynamics for $t>t_{s}$ is described by the following
master equation
\begin{subequations}
\label{ME-kappa-zero}%
\begin{align}
&  ~~~~~\dot{\rho}\left(  t\right)  =-i\left[  \widetilde{H}_{S}\left(  t
\right)  ,\rho\left(  t\right)  \right]  +L\left[  \rho\left(  t\right)
\right]  ,\\
&  ~~\widetilde{H}_{S}\left(  t\right)  =\widetilde{\epsilon}_{+}\left(
t\right)  A_{+}^{\dagger}A_{+}+\epsilon_{-}A_{-}^{\dagger}A_{-},\\
&  L\left[  \rho\left(  t\right)  \right]  =\widetilde{\kappa}\left(
t\right)  \left[  2A_{+}^{\dagger}\rho\left(  t\right)  A_{+}-A_{+}%
A_{+}^{\dagger}\rho\left(  t\right)  -\rho\left(  t\right)  A_{+}%
A_{+}^{\dagger}\right]  .
\end{align}
\end{subequations}
Then the states $|+\rangle$ and $|\mathsf{d}%
\rangle$ become possible dynamically stabilized DF states,
since $A_{+}^{\dag}|+\rangle=0$ and also
$A_{+}^{\dag}|\mathsf{d}\rangle=0$ leads to the vanishing of the nonunitary part $L[\rho(t)]$ after
$t>t_s$.
Therefore, the Markovian DFS criterion of
Eq.~(\ref{criterion-1-for-DFS}) still is both necessary and sufficient
for a time-local DFS when only one decoherence rate is turned on.

On the other hand, if we apply a negative bias to the electron reservoir, e.g.
$\mu=-10\Gamma$, as shown by the black curves in Fig. \ref{fig_kappa}, we find
that for both weakly and strongly non-Markovian dynamics, the decoherence rate
$\widetilde{\kappa}(t)$ goes to zero very quickly, while $\kappa(t)$ either
reaches a nonzero steady value (in the weakly non-Markovian regime), or keeps
jumping from a positive value to a negative value for all times (in the
strongly non-Markovian regime). Then the master equation for $t>t_s$ is effectively given
by
\begin{subequations}
\label{ME-tkappa-zero}%
\begin{align}
&~~~~~\dot{\rho}\left(  t\right) =-i\left[  \widetilde{H}_{S}\left(  t\right)
,\rho\left(  t\right)  \right]  +L\left[  \rho\left(  t\right)  \right]  ,\\
&~~\widetilde{H}_{S}\left(  t\right) =\widetilde{\epsilon}_{+}\left(
t\right)  A_{+}^{\dagger}A_{+}+\epsilon_{-}A_{-}^{\dagger}A_{-},\\
&L\left[  \rho\left(  t\right)  \right]=\kappa\left(  t\right)  \left[
2A_{+}\rho\left(  t\right)  A_{+}^{\dagger}-A_{+}^{\dagger}A_{+}\rho\left(
t\right)  -\rho\left(  t\right)  A_{+}^{\dagger}A_{+}\right]  .
\end{align}
\end{subequations}
Again, the DFS criterion Eq.~(\ref{criterion-1-for-DFS}) is then both
necessary and sufficient for DF states.
Since $A_+|-\rangle=0$ and $A_+|\mathsf{v}\rangle=0$, then we obtain that
$|-\rangle$ and $|\mathsf{v}\rangle$ are possible dynamically
stabilized DF states.

In summary, the above results show that the vanishing of one of the
decoherence rates implies that
$\left\{  |+\rangle
,|\mathsf{d}\rangle\right\}$ of $\left\{  |-\rangle
,|\mathsf{v}\rangle\right\}$ become dynamically stabilized DF states.

\section{Physical realization for dynamically stabilized DF states}
\label{chap-V}

In the following, by examining the general exact solution of
master Eq.~(\ref{exact-master-equation}),
we will prove that all pure dynamically stabilized DF states
$\{|+\rangle,|\mathsf{d}\rangle\}$ and $\{|-\rangle
,|\mathsf{v}\rangle\}$ predicted above are physically
realizable through the decoherence process given by the same master
equation.
We also give the exact conditions for generating these DF states in
terms of the initial state, and a condition on the Green's functions
$u(t)$ and $v(t)$.

First, by solving (\ref{exact-master-equation}), we can exactly give the elements
of $\rho\left(  t\right)  $ in Eq. (\ref{general-rhot}) in terms of the initial
$\rho\left(  t_{0}\right)  $ and the functions of $u\left(  t\right)  $ and
$v\left(  t\right)  $ as
\begin{align}
&  \rho_{\mathsf{vv}}\left(  t\right)  =[ 1-v\left(  t\right)  ]
\rho_{\mathsf{vv}}\left(  t_{0}\right)  +[ 1-v\left(  t\right)  -\left\vert
u\left(  t\right)  \right\vert ^{2}] \rho_{++}\left(  t_{0}\right)
,\nonumber\\
&  \rho_{++}\left(  t\right)  =v\left(  t\right)  \rho_{\mathsf{vv}}\left(
t_{0}\right)  +[ v\left(  t\right)  +\left\vert u\left(  t\right)  \right\vert
^{2}] \rho_{++}\left(  t_{0}\right)  ,\nonumber\\
&  \rho_{+-}\left(  t\right)  =u\left(  t\right)  e^{i\epsilon_{-}t}\rho
_{+-}\left(  t_{0}\right)  ,\nonumber\\
&  \rho_{--}\left(  t\right)  =[ 1-v\left(  t\right)  ] \rho_{--}\left(
t_{0}\right)  +[ 1-v\left(  t\right)  -\left\vert u\left(  t\right)
\right\vert ^{2}] \rho_{\mathsf{dd}}\left(  t_{0}\right)  ,\nonumber\\
&  \rho_{\mathsf{dd}}\left(  t\right)  =v\left(  t\right)  \rho_{--}\left(
t_{0}\right)  +[ v\left(  t\right)  +\left\vert u\left(  t\right)  \right\vert
^{2}] \rho_{\mathsf{dd}}\left(  t_{0}\right). \label{element-rhot-two-dots}
\end{align}
This expression for $\rho\left(  t\right)  $
is valid for an arbitrary spectral density of the reservoir. This solution
confirms the fact that $\mathcal{H}_{+}=\left\{  |\mathsf{v}\rangle
,|+\rangle\right\}  $ and $\mathcal{H}_{-}=\left\{  |-\rangle,|\mathsf{d}%
\rangle\right\}  $ are two independent closed subspaces. For any initial
state in the subspace $\mathcal{H}_{+}$ $\left(  \mathcal{H}_{-}\right)  $,
the system will be dynamically stabilized in this subspace.

The general solution (\ref{element-rhot-two-dots}) shows that if $v\left(  t\right)  =1$ is
satisfied when $t>t_s$, the initial vacuum state $|\mathsf{v}\rangle$
converges to
the stabilized state $|+\rangle$, while if $v\left(  t\right)  +\left\vert
u\left(  t\right)  \right\vert ^{2}=1$ is reached when $t>t_s$, the initial
singly occupied state
$|+\rangle$ converges to
the dynamically stabilized DF state
$|+\rangle$. That is, the same dynamically stabilized DF state
$|+\rangle$ can be generated from different initial states in the same
subspace of
$\mathcal{H}_{+}$ under
different stabilization conditions for $u(t)$ and
$v(t)$. Similarly, one can
obtain the remaining dynamically stabilized DF states $|\mathsf{v}\rangle$,
$|-\rangle$ and $|\mathsf{d}\rangle$ from different initial states under different stabilization condition, as shown
in Table \ref{table}.

\begin{table}[h]
\caption{The dynamically stabilized DF states for different choices of the
initial state of the system, plus the different stabilization conditions.}%
\label{table}
\begin{center}%
\begin{tabular}
[c]{|c|c|c|}\hline
initial state & stabilization condition & DF state\\\hline
$|\mathsf{v}\rangle$ & $v\left(  t\right)  =0$ (or $1$) & $|\mathsf{v}\rangle$
(or $|+\rangle$)\\\hline
$|+\rangle$ & $v\left(  t\right)  +\left\vert u\left(  t\right)  \right\vert
^{2}=0$ (or $1$) & $|\mathsf{v}\rangle$ (or $|+\rangle$)\\\hline
$|-\rangle$ & $v\left(  t\right)  =0$ (or $1$) & $|-\rangle$ (or $|$d$\rangle
$)\\\hline
$|\mathsf{d}\rangle$ & $v\left(  t\right)  +\left\vert u\left(  t\right)
\right\vert ^{2}=0$ (or $1$) & $|-\rangle$ (or $|$d$\rangle$)\\\hline
\end{tabular}
\newline\ \newline
\end{center}
\label{table}%
\end{table}

Besides the initial states listed in Table \ref{table}, another more general
initial pure state is a superposition of $|+\rangle$
and $|-\rangle$, namely $|\Phi\rangle=\alpha|+\rangle+ \beta|-\rangle$
with $|\alpha|^{2}+|\beta|^{2}=1$. In this case, as one can easily check from
Eq.~(\ref{element-rhot-two-dots}), the resulting state becomes
\begin{align}
&  \rho_{\mathsf{vv}}\left(  t\right)  = [ 1-v\left(  t\right)  -\left\vert
u\left(  t\right)  \right\vert ^{2}] |\alpha|^{2} ,\nonumber\\
&  \rho_{++}\left(  t\right)  = [v\left(  t\right)  +\left\vert u\left(
t\right)  \right\vert ^{2}] |\alpha|^{2} ,\nonumber\\
&  \rho_{+-}\left(  t\right)  =u\left(  t\right)  e^{i\epsilon_{-}t}%
\ \alpha\beta^{*} ,\nonumber\\
&  \rho_{--}\left(  t\right)  =\left[  1-v\left(  t\right)  \right]
|\beta|^{2} ,\nonumber\\
&  \rho_{\mathsf{dd}}\left(  t\right)  =v\left(  t\right)  |\beta|^{2} .
\label{element-rhot-two-dots-1}%
\end{align}
In the case that both $\alpha$ and $\beta$ are nonzero,
one has to have $v(t)=0$ and
$|u(t)|^{2}=1$ to generate a stabilized (pure) DF state $|\Phi\rangle$. This
is impossible unless the system
is totally decoupled from the environment from the beginning. In other words,
except for a stabilized mixed state, no stabilized DF state can be obtained.
As a conclusion, Table \ref{table} lists all the possible pure stabilized
DF states in
this system. The present results confirm the statement in the last
section that the
only possible pure dynamically stabilized DF states are $|v\rangle,|\pm\rangle$
 and $|d\rangle$.

\begin{figure}[ptb]
\begin{center}
\includegraphics[height=7.3cm,width=9cm]{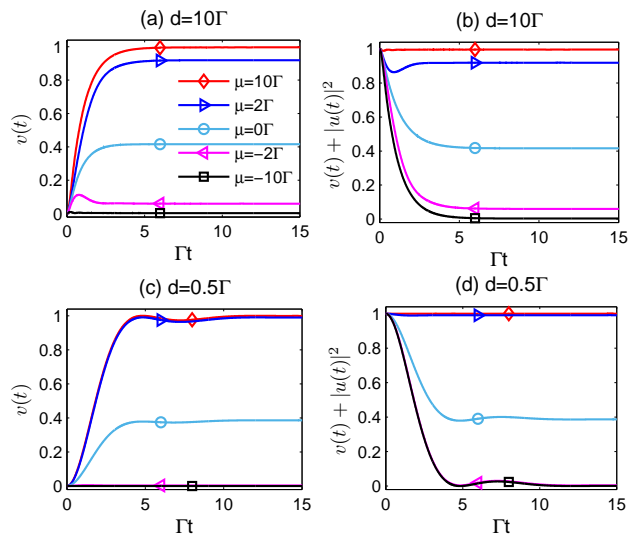}
\end{center}
\caption{The exact solution of $v(t)$ and $v(t)+|u(t)|^{2}$ for an electron
reservoir with varying the external bias voltage $\mu=eV$. (a)-(b) for the
weakly non-Markovian dynamics (with $d=10\Gamma$); (c)-(d) for the strongly
non-Markovian dynamics (with $d=0.5\Gamma$ ). Here we set $\epsilon_{0}
=0.2\Gamma$, ${K_{B}}T=0.3\Gamma$. By applying a large positive (or negative)
bias voltage, the steady value of $v(t)$ and $v(t)+|u(t)|^{2}$ will approach 1
(or 0). }%
\label{fig_vt}%
\end{figure}

For the Lorentz spectral density, we display the functions of $v(t)$ and $v(t)+|u(t)|^2$ in Fig.~\ref{fig_vt}.
Interestingly, by comparing Fig.~\ref{fig_vt} (a)-(b) and (c)-(d), we find
that in the strongly non-Markovian case ($d=0.5\Gamma$), applying a relatively
small bias voltage (for example $\left\vert \mu\right\vert =2\Gamma$) makes
the steady values of $v\left(  t\right)  $ and $v\left(  t\right)  +\left\vert
u\left(  t\right)  \right\vert ^{2}$ quickly approach $1$ or $0$, whereas this
is not the case in the weakly non-Markovian regime ($d=10\Gamma$), where
$v(t)\to0.9$ for $\mu=2\Gamma$ (and $v(t)\to0.1$ for $\mu=-2\Gamma$). This
indicates that the back-flow of information from the reservoir,
due to a strongly non-Markovian memory of the reservoir, helps the
stabilization of the system in the states of $\{|+\rangle,|\mathsf{d}%
\rangle\}$ or $\{|\mathsf{v}\rangle,|-\rangle\}$. This facilitates the
physical realization of the dynamically stabilized DF states.

Furthermore, we can now prove that under the
stabilization conditions listed in Table \ref{table},
one of the decoherence rates vanishes when $t>t_s$. In fact,
$\kappa(t)$ and $\widetilde{\kappa}(t)$ in Eq.~(\ref{decohering-coefficients})
can be further simplified as
\begin{subequations}
\begin{align}
\kappa\left(  t\right)   &  =\frac{\left\vert u\left(  t\right)  \right\vert
^{2}}{2}\frac{d}{dt}\frac{1-v\left(  t\right)  }{\left\vert u\left(  t\right)
\right\vert ^{2}},\\
\widetilde{\kappa}\left(  t\right)   &  =\frac{\left\vert u\left(  t\right)
\right\vert ^{2}}{2}\frac{d}{dt}\frac{v\left(  t\right)  }{\left\vert u\left(
t\right)  \right\vert ^{2}}.
\end{align}
\end{subequations}
It clearly shows that the condition $v(t)=1$ (or ($v(t)+|u(t)|^2=1$) implies
$\kappa(t)=0$, and the condition $v(t)=0$ (or $v(t)+|u(t)|^2=0$)
indicates $\widetilde{\kappa}(t)=0$, namely, one decoherence rate is
turned off for $t>t_s$. In this case, the Markovian DFS criterion still
provides a necessary and sufficient condition for generating dynamically
stabilized DF states. However, for the non-Markovian dynamics, we
have shown in Table \ref{table} that which final state is realized
depends both on the initial state and the details of the
dynamics given by the Green functions $u(t)$ and $v(t)$.

\section{Comparison to the Born-Markov dynamics}

In the previous Sections, using the exact master equation, we have shown the
generation of DF states for a system coupled to a non-Markovian reservoir.
For comparison, in the following, we discuss the corresponding results in
terms of the BM master equation.

The BM dynamics usually corresponds to the case where the coupling strength
between the system and the electron reservoir is very weak, and the
characteristic correlation time of the electron reservoir is sufficiently
shorter than that of the system. In such a case, the electron reservoir has no
memory effect on the evolution of the system. Then the solution of $u\left(
t\right)  $, $v\left(  t\right)  $ are reduced to
\cite{TuZhang08,TuZhang09,JinZhang10}
\begin{align}
u_{\text{BM}}\left(  t\right)   &  =e^{ - i\widetilde{\epsilon}_{+}t-\frac{1}%
{2}J\left(  \epsilon_{+}\right)  t} ,\nonumber\\
v_{\text{BM}}\left(  t\right)   &  =\big[ 1-e^{-J\left(  \epsilon_{+}\right)
t}\big] f\left(  \epsilon_{+}\right)  , \label{uv-BM}%
\end{align}
where $\widetilde{\epsilon}_+=\epsilon_{+}+\left(  \delta\epsilon_{+}\right)
_{\text{BM}}$ with the energy shift $\left(  \delta\epsilon_{+}\right)
_{\text{BM}}=\mathcal{P}\int_{-\infty}^{+\infty}\frac{d\omega}{2\pi}%
\frac{\Gamma\left(  \omega\right)  }{\omega-\epsilon_{+}}$. Note that for the
wide-band limit, we simply have $\widetilde{\epsilon}_+=\epsilon_{+}$ and
$J(\epsilon_{+})=\Gamma$. Substituting Eq.~(\ref{uv-BM}) into
Eq.~(\ref{decohering-coefficients}), we obtain the constant rates
\begin{align}
\kappa\left(  t\right)   &  =\frac{1}{2}J\left(  \epsilon_{+}\right)  \left[
1-f\left(  \epsilon_{+}\right)  \right]  ,\nonumber\\
\widetilde{\kappa}\left(  t\right)   &  =\frac{1}{2}J\left(  \epsilon
_{+}\right)  f\left(  \epsilon_{+}\right)  . \label{BM dr}%
\end{align}
Here $f\left(  \epsilon_{+}\right)  $ is the fermion distribution function of
the electron reservoir at the frequency $\epsilon_{+}$, i.e.,
\begin{equation}
f\left(  \epsilon_{+}\right)  =\frac{1}{e^{\left(  \epsilon_{+}-\mu\right)
/K_{B}T}+1}.
\end{equation}
Thus the exact master equation is reduced to the BM master equation, where the
decoherence rates are time-independent. This gives the standard Lindblad form
for the Markovian dynamics. Based on the general DFS criterion of
Eq.~(\ref{criterion-1-for-DFS}), we see that there is in general no DFS
for this system in the BM limit.

However, if we apply a large positive bias $\mu=eV$ such that, $\left(
\mu-\epsilon_{+}\right)  /k_{B}T\gg1$, then $f\left(  \epsilon_{+}\right)
\rightarrow1$, which leads to $\kappa  \rightarrow 0$. $\widetilde
{H}_{S}$ leaves the states $|+\rangle$ and  $|$d$\rangle$ invariant
during the time evolution. The relation
$A_{+}^{\dagger}|+\rangle=0$ (or $A_{+}^{\dagger}|$d$\rangle=0$) guarantees
$L\left[  \rho\left(  t\right)
\right]  =0$ for these states. The states $\left\{
|+\rangle,|\text{d}\rangle\right\}  $
are therefore DF states in the BM limit under large positive
bias. Likewise, applying a negative
large bias
to the electron reservoir such that $\left(  \epsilon_{+}-\mu\right)
/K_{B}T\gg1$, then $f\left(  \epsilon_{+}\right)  \rightarrow0$ and
$\widetilde{\kappa}(t)\rightarrow0$. In this case, the states $\left\{
|\text{v}\rangle,|-\rangle\right\}  $ are the DF states in the BM limit.

In conclusion, the states $\left\{  |+\rangle,|\text{d} \rangle\right\}  $
and $\left\{  |\text{v}\rangle,|-\rangle\right\}  $ are also DF in the
BM limit if one of the decoherence rates is switched off by
properly tuning the bias voltage on the electron reservoir. This result is
consistent with the result in the non-Markovian case discussed above. The
apparent
difference is that the DF states in the BM limit seem to exist without the
dynamical stabilization processes.
However, this difference is not crucial in reality. As is well-known
\cite{Carmichael93,Breuer07}, the BM master equation with the constant
decoherence rates, Eq.~(\ref{BM dr}), is derived under the condition $t\gg
\tau_{r}$ where $\tau_{r}$ is the characteristic time of the reservoir
\cite{Carmichael93}. In other words, a stabilization time scale $t_{s}$
has implicitly been used in deriving the  BM master equation, such
that the decoherence rates become time-independent
for $t>t_{s} \gg\tau_{r}$ . Therefore,
the concept of the dynamically stabilized DF states, based on the exact
non-Markovian master equation, gives the generalized picture of
DFS. It contains
the Markovian DFS as a special case.\\

\section{Conclusion}

In summary, we have investigated the DFS of a non-Markovian fermionic open
systems, based on an exact non-Markovian master
equation developed recently. The master equation has a nonunitary
term of the standard Lindblad form, but the corresponding
decoherence rates are
time-dependent and local in time. They are determined microscopically and
nonperturbatively from the Schwinger-Keldysh nonequilibrium Green
functions and fully account for the non-Markovian memory effect.

As concrete example we have studied  a fermionic system with two
degenerate energy levels coupled identically to a fermionic
reservoir.  We find that the
whole Hilbert space is split into two closed subspaces. For any initial state
in one of the subspaces, the system will remain in this subspace
forever.
By diagonalizing the full Lindblad operator we found that physical DF
states exist if and only if one of the two relevant decoherence rates
switches itself off dynamically. Such a situation can be achieved physically.
Two of the DF states are coherent
superpositions with an arbitrary relative phase between the two original
energy levels, which may be of physical interest for quantum computation.

Which DF state is reached as result of the dissipative dynamics
depends both on the initial state and the details of the dynamics, as
expressed by the time-dependent non-equilibrium Green's function. We
show this explicitly by
solving exactly the non-Markovian master equation.
Interestingly, we find that the strongly non-Markovian memory can help to
stabilize the DF states compared to the Markovian case.

\begin{acknowledgements}
This work is supported by the National Science
Council of ROC under Contract No. NSC-99-2112-M-006-008-MY3, the National
Center for Theoretical Science of Taiwan.
\end{acknowledgements}

\end{document}